\documentclass[twocolumn,aps,showpacs,superscriptaddress,tightenlines,a4paper]{revtex4}

\usepackage{amssymb}
\usepackage{amsmath}
\usepackage{bbold}
\usepackage{mathrsfs}
\usepackage{paralist}
\usepackage{graphicx}
\usepackage{xcolor}

\newcommand{\ii}{{\rm i}}
\newcommand{\e}{{\rm e}}

\begin{document}
%
%
	\title{Active absorption of electromagnetic pulses in a cavity}
%
%
    \author{S. A. R. Horsley}
    \affiliation{Department of Physics and Astronomy, University of Exeter,
Stocker Road, Exeter, EX4 4QL}
    \email{s.horsley@exeter.ac.uk}
    \author{R. N. Foster}
    \affiliation{Queen Mary University of London,
Mile End Rd, London, E1 4NS}
    \author{T. Tyc}
    \affiliation{Department of Theoretical Physics and Astrophysics, Masaryk University,
Kotl\'a\v{r}sk\'a 2, 61137 Brno, Czech Republic}
    \author{T. G. Philbin}
    \affiliation{Department of Physics and Astronomy, University of Exeter,
Stocker Road, Exeter, EX4 4QL}
    
%
%
    \begin{abstract}
    	We show that a pulse of electromagnetic radiation launched into a cavity can be completely absorbed into an infinitesimal region of space, provided one has a high degree of control over the current flowing through this region.  We work out explicit examples of this effect in a cubic cavity and a cylindrical one, and experimentally demonstrate the effect in the microwave regime.
    \end{abstract}
%
%
    \maketitle
%
%
%
%
	\section{Introduction}
	\par
	There is a fundamental limit to the amount of radiation that can be absorbed by a passive system of a given size.  For instance one form of the Rozanov limit~\cite{rozanov2000} relates the thickness of a planar medium to the bandwidth over which it can be an efficient absorber.  This limit has its origin in the Kramers--Kronig relations~\cite{volume5}, and generally holds for any such \emph{passive} system.  However, this does not apply to an active system of currents; based on some knowledge of the radiation that one is trying to absorb, an active system can eliminate a polychromatic field, even though it might only occupy a small region of space.  This was demonstrated by de Rosny and Fink~\cite{derosny2002} through placing an active `drain' (a current source driven in reverse) in the focus of an incoming sound wave, showing that the wave could be concentrated and absorbed within a region of space much less than a wavelength in size.
	
	The use of an active current element to concentrate the electromagnetic field was recently the subject of debate in the context of imaging~\cite{leonhardt2009,blaikie2010,kinsler2011,ma2011,tyc2011,teruel2012,tyc2014}, where it was claimed that the sub-wavelength scale of the field around an active drain could be used to resolve small features of a distant object within the Maxwell fish--eye lens.  Although it now seems unlikely that a useful imaging device can be made in this way~\cite{teruel2012}, the device proposed in~\cite{leonhardt2009} has other interesting properties.  In particular, Tyc and Danner~\cite{tyc2012} have shown that absolute optical instruments~\footnote{For an absolute optical instrument, there is a region of space in which any point A has a sharp (stigmatic) image B; this means that infinitely many rays from A get to B.  In the case of the Maxwell fish–eye lens~\cite{born2002}, A can be any point in the device.} tend to have a nearly uniform spacing of eigenfrequencies.  Such a spectrum allows one to emit a pulse from a point and then later absorb all the radiation in the device through emitting a second pulse through the same point.  This is similar in spirit to the work of Fink and co--workers~\cite{derosny2002} except that rather than using time reversal of the field to focus it onto the drain, the properties of the spectrum of the device ensure this instead, suggesting a close connection to the phenomenon of wave--packet revival~\cite{robinett2004}.
	
	In this work we demonstrate that the effect noticed by Tyc and Danner in absolute optical instruments can also be observed in an empty cavity.  We show that within a cubic cavity one can emit a pulse from a point, and then completely reabsorb it at a later time through emitting a second pulse from the same point, or one related by mirror symmetry.  Finally, we show an experimental demonstration of this effect.
%
%
	\section{Electromagnetic radiation from a source in a cavity}
	\par
	We begin by recalling the behaviour of a source of electromagnetic radiation in a cavity with perfectly conducting walls.   The radiation generated from a time dependent current density \(\boldsymbol{j}\) satisfies the inhomogeneous electromagnetic wave equation
	\begin{equation}	\boldsymbol{\nabla}\boldsymbol{\times}\boldsymbol{\nabla}\boldsymbol{\times}\boldsymbol{E}+\frac{1}{c^{2}}\frac{\partial^{2}\boldsymbol{E}}{\partial t^{2}}=-\mu_{0}\frac{\partial\boldsymbol{j}}{\partial t}.
	\end{equation}
	This has the general solution
	\begin{equation}
		\boldsymbol{E}(\boldsymbol{x},t)=\ii\mu_{0}\int_{-\infty}^{\infty}\frac{d\omega}{2\pi}\,\omega \e^{-\ii\omega t}\int_{V} \boldsymbol{G}(\boldsymbol{x},\boldsymbol{x}^{\prime},\omega)\boldsymbol{\cdot}\boldsymbol{j}(\boldsymbol{x}^{\prime},\omega)d^{3}\boldsymbol{x}^{\prime}\label{electric-field}
	\end{equation}
	where \(V\) is the volume of space occupied by the cavity, and the electromagnetic Green function \(\boldsymbol{G}\) (a dyadic) satisfies
	\begin{equation}
		\boldsymbol{\nabla}\boldsymbol{\times}\boldsymbol{\nabla}\boldsymbol{\times}\boldsymbol{G}(\boldsymbol{x},\boldsymbol{x}',\omega)-\frac{\omega^{2}}{c^{2}}\boldsymbol{G}(\boldsymbol{x},\boldsymbol{x}',\omega)=\boldsymbol{\mathbb{1}}_{3}\delta^{(3)}(\boldsymbol{x}-\boldsymbol{x}').
	\end{equation}
	The retarded Green function can be expanded in terms of the eigenmodes of the system, which in the case of a cavity occur at discrete frequencies \(\omega_{n}\),
	\begin{multline}
		\boldsymbol{G}(\boldsymbol{x},\boldsymbol{x}^{\prime},\omega)=-c^{2}\sum_{n}\frac{\boldsymbol{E}_{n}(\boldsymbol{x})\boldsymbol{\otimes}\boldsymbol{E}_{n}(\boldsymbol{x}^{\prime})}{(\omega+\ii\eta)^{2}-\omega_{n}^{2}}\\
		-\frac{c^{2}}{(\omega+\ii\eta)^{2}}\boldsymbol{\delta}_{\parallel}(\boldsymbol{x}-\boldsymbol{x}^{\prime})\label{gf},
	\end{multline}
	where `\(\boldsymbol{\otimes}\)' indicates a tensor product.  There is also a sum over the polarization degree of freedom which is implicit in (\ref{gf}), the details of which can be found in appendix~\ref{apa}.  In (\ref{gf}) \(\eta\) is an infinitesimal positive number which we take to zero at the end of every calculation, and \(\boldsymbol{\delta}_{\parallel}(\boldsymbol{x}-\boldsymbol{x}_{0})\) is the longitudinal part of the delta function~\cite{craig1984}, which is shown only for completeness and plays no role in the rest of this calculation.  The \(\boldsymbol{E}_{n}(\boldsymbol{x})\) are the eigenfunctions of the cavity in the absence of any source,
	\[
		\boldsymbol{\nabla}\boldsymbol{\times}\boldsymbol{\nabla}\boldsymbol{\times}\boldsymbol{E}_{n}-\frac{\omega_{n}^{2}}{c^{2}}\boldsymbol{E}_{n}=0,
	\]
	normalized such that,
	\begin{equation}
		\int_{V}\boldsymbol{E}_{n}(\boldsymbol{x})\boldsymbol{\cdot}\boldsymbol{E}_{m}(\boldsymbol{x})d^{3}\boldsymbol{x}=\delta_{nm},\label{orthonormal}
	\end{equation}
	It is assumed that the eigenmodes \(\boldsymbol{E}_{n}\) are real valued vector fields.
	\par
	Suppose there is a point--like current \(\boldsymbol{j}_{0}(\boldsymbol{x},t)=\delta^{(3)}(\boldsymbol{x}-\boldsymbol{x}_{0})\boldsymbol{\mathcal{J}}_{0}(t)\) located within a cavity, switched on for a time interval \(\Delta t_{0}\).  Such a current has the following frequency domain representation,
	\begin{align}
		\boldsymbol{j}_{0}(\boldsymbol{x},\omega)&=\delta^{(3)}(\boldsymbol{x}-\boldsymbol{x}_{0})\int_{t_{0}-\Delta t_{0}/2}^{t_{0}+\Delta t_{0}/2}\boldsymbol{\mathcal{J}}_{0}(t)\e^{\ii\omega t}d t\nonumber\\
		&=\delta^{(3)}(\boldsymbol{x}-\boldsymbol{x}_{0})\boldsymbol{\mathcal{J}}_{0}(\omega).\label{jw}
	\end{align}
	We note that in order for the radiation source to remain uncharged, the net charge transferred to the source must be zero
	\begin{equation}
	\int_{-\infty}^{\infty}\boldsymbol{\mathcal{J}}_{0}(t)d t=\boldsymbol{\mathcal{J}}_{0}(\omega=0)=0\label{Jw0},
	\end{equation}
	This means that the Fourier component of the source at zero frequency must vanish.
	\par
	When computed from expression (\ref{electric-field}), the electric field in the time domain is given as an integral over frequency.  For times prior to the start of the pulse (\(t<t_{0}-\Delta t_{0}/2\)), the integral over \(\omega\) can be replaced with a contour integral closed in the \emph{upper} half frequency plane.  The Green function (\ref{gf}) is analytic in the upper half plane and this integral is zero.  We have thus established the obvious fact that the field in the cavity is zero before the current is turned on
	\begin{equation}
		\boldsymbol{E}(\boldsymbol{x},t<t_{0}-\Delta t_{0}/2)=0.
	\end{equation}
	Meanwhile, when \(t>t_{0}+\Delta t_{0}/2\), the integral over frequency in (\ref{electric-field}) may be replaced with a contour integral closed in the \emph{lower} half frequency plane.  Within this contour, the Green function (\ref{gf}) has poles at \(\omega=\pm\omega_{n}-\ii\eta\), and an application of the residue theorem along with (\ref{Jw0}) yields,
	\begin{multline}
		\boldsymbol{E}(\boldsymbol{x},t>t_{0}+\Delta t_{0})=-\frac{1}{\epsilon_{0}}\sum_{n}\boldsymbol{E}_{n}(\boldsymbol{x})\\
		\times\boldsymbol{E}_{n}(\boldsymbol{x}_{0})\boldsymbol{\cdot}\text{Re}[\boldsymbol{\mathcal{J}}_{0}(\omega_{n})\e^{-\ii\omega_{n} t}]\label{pulse}.
	\end{multline}
	To obtain (\ref{pulse}), we assumed that the current takes a real value in the time domain so that \(\mathcal{J}_{0}(\omega)=\mathcal{J}^{\star}_{0}(-\omega)\).  The pole at zero frequency, evident in the longitudinal part of the Green function (\ref{gf}) does not contribute, because we have assumed condition (\ref{Jw0}).  Equation (\ref{pulse}) means that after the current has been switched off (\(t>t_{0}+\Delta t_{0}/2\)) the electric field reduces to a sum over the eigenmodes of the cavity, each weighted by the corresponding Fourier component of the source \(\boldsymbol{\mathcal{J}}_{0}(\omega_{n})\).
	
	Due to the dependence of (\ref{pulse}) on the Fourier amplitude of the current at discrete eigenfrequencies of the cavity, two different current pulses can produce the same final field; they only have to interpolate the same value at the eigenfrequencies.  This freedom allows us to emit two pulses, with the second pulse serving to completely absorb the first.  In the next section we work out the details of this phenomenon.
%
%
	
	\section{Absorbing radiation through emitting radiation\label{abs-sec}}
	\par
	For an ideal cavity with perfectly reflecting walls, the total energy in the electromagnetic field will remain constant after the current pulse \(\boldsymbol{\mathcal{J}}_{0}\) has finished.  But suppose we want to reduce the energy within the cavity to zero through emitting a second pulse through \(\boldsymbol{x}_{1}\).  What kind of secondary pulse would be required?
	\par
	In the following discussion we consider the case when \(\boldsymbol{x}_{0}\) and \(\boldsymbol{x}_{1}\) are different points, but a very similar analysis applies when they are the same point~\cite{tyc2012}.  If a second current, \(\boldsymbol{\mathcal{J}}_{1}(t)\) passes through \(\boldsymbol{x}_{1}\) during a time interval \(\Delta t_{1}\), then the generalization of (\ref{pulse}) for times later than \(t_{1}+\Delta t_{1}/2\) is,
	\begin{multline}
		\boldsymbol{E}(\boldsymbol{x},t>t_{1}+\Delta t_{1})=-\frac{1}{\epsilon_{0}}\sum_{n}\boldsymbol{E}_{n}(\boldsymbol{x})\\
		\times\text{Re}\left\{[\boldsymbol{E}_{n}(\boldsymbol{x}_{0})\boldsymbol{\cdot}\boldsymbol{\mathcal{J}}_{0}(\omega_{n})+\boldsymbol{E}_{n}(\boldsymbol{x}_{1})\boldsymbol{\cdot}\boldsymbol{\mathcal{J}}_{1}(\omega_{n})]\e^{-\ii\omega_{n} t}\right\}.\label{abs-field}
	\end{multline}
	If the second pulse completely absorbs the first, then (\ref{abs-field}) must equal zero.  The only way for this to happen is when the Fourier amplitudes of the two current pulses are related by
	\begin{equation}
		\boldsymbol{E}_{n}(\boldsymbol{x}_{0})\boldsymbol{\cdot}\boldsymbol{\mathcal{J}}_{0}(\omega_{n})=-\boldsymbol{E}_{n}(\boldsymbol{x}_{1})\boldsymbol{\cdot}\boldsymbol{\mathcal{J}}_{1}(\omega_{n}).\label{current-2}
	\end{equation}
	It is non--trivial to satisfy this for all possible pulses, and usually impossible without some restriction on \(\boldsymbol{x}_{1}\).  In the next section we construct expressions for \(\boldsymbol{\mathcal{J}}_{1}(t)\) in two example cavities.
	\par
	To quantify the degree of absorption we use the energy of the radiation contained in the cavity as a function of time, \(\mathscr{E}(t)\)
	\begin{align}
		\mathscr{E}(t)&=\frac{\epsilon_{0}}{2}\int_{V}d^{3}\boldsymbol{x}\left[\boldsymbol{E}^{2}+c^{2}\boldsymbol{B}^{2}\right]\nonumber\\
				&=\frac{\epsilon_{0}}{2}\int_{V}d^{3}\boldsymbol{x}\left[\dot{\boldsymbol{A}}^{2}+c^{2}\boldsymbol{\nabla}\boldsymbol{\times}\boldsymbol{A}^{2}+\left(\boldsymbol{\nabla}\varphi\right)^{2}\right]
	\end{align}
	where the fields have been written in terms of the scalar and vector potentials, \(\boldsymbol{E}=-\boldsymbol{\nabla}\varphi-\dot{\boldsymbol{A}}\) and \(\boldsymbol{B}=\boldsymbol{\nabla}\boldsymbol{\times}\boldsymbol{A}\), and we imposed the Coulomb gauge \(\boldsymbol{\nabla}\boldsymbol{\cdot}\boldsymbol{A}=0\)~\cite{jackson2002}.  Expanding the vector potential in terms of the eigenfunctions of the cavity
	\begin{equation}
		\boldsymbol{A}(\boldsymbol{x},t)=\sum_{n}\mathcal{C}_{n}(t)\boldsymbol{E}_{n}(\boldsymbol{x})\label{A-exp}
	\end{equation}
	and applying (\ref{orthonormal}), the energy in the cavity becomes
	\begin{equation}
		\mathscr{E}_{R}(t)=\frac{\epsilon_{0}}{2}\sum_{n}\left[\dot{\mathcal{C}}_{n}(t)^{2}+\omega_{n}^{2}\mathcal{C}_{n}(t)^{2}\right]\label{cavity-energy}
	\end{equation}
	where the contribution due to the scalar potential has been dropped (in any case (\ref{Jw0}) ensures this term is zero before and after the source acts).  The expansion coefficients appearing in the energy in (\ref{cavity-energy}) can be found from an examination of (\ref{electric-field}) and (\ref{gf}), resulting in
	\begin{multline}
		\mathcal{C}_{n}(t)=\int_{-\infty}^{\infty}d\omega\,\frac{\e^{-\ii\omega t}}{2\pi\epsilon_{0}(\omega_{n}^{2}-(\omega+\ii\eta)^{2})}[\boldsymbol{E}_{n}(\boldsymbol{x}_{0})\boldsymbol{\cdot}\boldsymbol{\mathcal{J}}_{0}(\omega)\\
		+\boldsymbol{E}_{n}(\boldsymbol{x}_{1})\boldsymbol{\cdot}\boldsymbol{\mathcal{J}}_{1}(\omega)]\label{expansion-coefficient}
	\end{multline}
	which is automatically real because the current is real in the time domain.  The energy in the cavity is calculated through inserting (\ref{expansion-coefficient}) into (\ref{cavity-energy}).
	
	In order to see the overall change of the energy with time we average \(\mathscr{E}_{R}(t)\) over a time window that is assumed long in comparison to the inverse frequencies within the pulse.  Writing \(\mathcal{C}_{n}(t)=(1/2)[c_{n}+c_{n}^{\star}]\), where \(c_{n}\) is equal to (\ref{expansion-coefficient}) with the integral extending over only positive frequencies, the averaging is equivalent to replacing \(\dot{\mathcal{C}}_{n}(t)^{2}\to\frac{1}{2}\left|\dot{c}_{n}(t)\right|^{2}\) and \(\mathcal{C}_{n}(t)^{2}\to\frac{1}{2}\left|c_{n}(t)\right|^{2}\).
%
%
	\subsection{Cubic cavity\label{cavsec}}
	%
	%
	\begin{figure}[h!]
		\begin{center}
		\includegraphics[width=8cm]{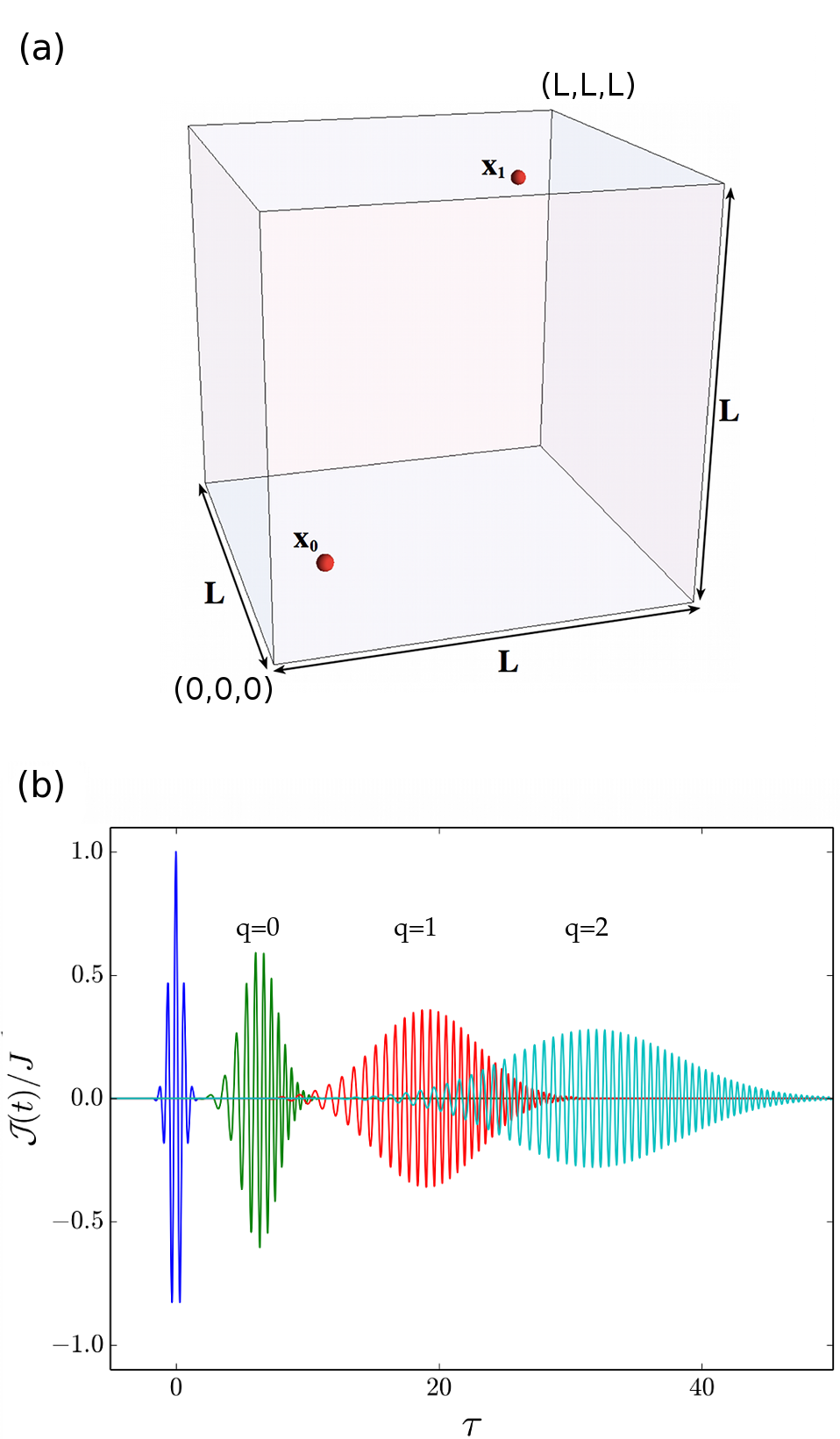}\\
		\end{center}
		\caption{(a) Two sources at opposite positions within a cubic cavity emit sequential pulses, the first of which, \(\mathcal{J}_{0}(t)\) is sent through \(\boldsymbol{x}_{0}\).  The second pulse, \(\mathcal{J}_{1}(t)\) is sent at a later time through \(\boldsymbol{x}_{1}\) and serves to reduce the field in the cavity to zero. (b) Normalized current pulses as a function of time (\(\tau=c t/L\) is a time variable normalized by the time taken to cross the cavity).  The initial pulse \(\mathcal{J}_{0}(t)\) is centred around \(\tau=0\), and \(\mathcal{J}_{1}(t)\) is shown for different choices of \(q\).  The initial pulse is given by (\ref{j0}), with the arbitrary values \(\sigma=2c/L\) and \(\omega_{0}=10c/L\).\label{set-up-box}}
	\end{figure} 
	\par
	In a cubic cavity there are analytical expressions for the eigenfunctions and eigenfrequencies, the details of which can be found in appendix~\ref{apa}.  If the side lengths are \(L\), the eigenfrequencies are given by
	\begin{equation}
		\omega_{n,m,p}=\frac{c\pi}{L}\sqrt{n^{2}+m^{2}+p^{2}}.\label{eigenfrequencies-1}
	\end{equation}
	where \(n,m,p\) are integers, one of which may be zero.   As we have seen, certainly the energy emitted at a point \(\boldsymbol{x}_0\) can be absorbed at the same point; in addition, as we will show now, it is possible to absorb it also at the opposite point (as shown in figure~\ref{set-up-box}a).
	
\begin{figure}[h!]
		\includegraphics[width=9.5cm]{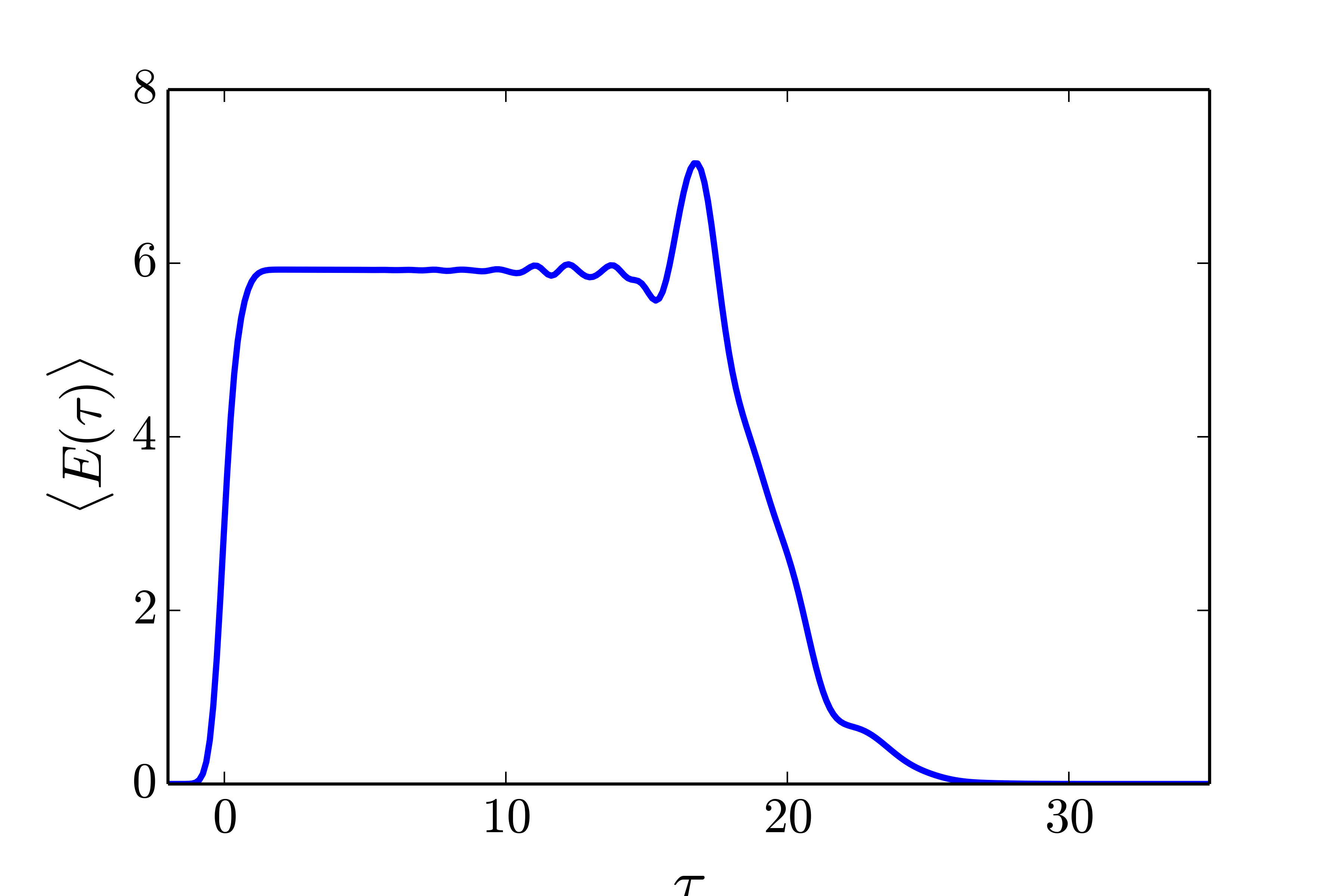}
		\caption{Cycle--averaged energy in the cavity in units of the characteristic energy \(\mu_{0}J^{2}/L\), \(\langle E(\tau)\rangle=\langle\mathscr{E}(\tau)\rangle L/\mu_{0}J^{2}\) computed from (\ref{cavity-energy}) as a function of time for the \(q=1\) case (see figure~\ref{set-up-box}).  After the second pulse has been emitted through \(\boldsymbol{x}_{1}\), the energy in the cavity is reduced to zero.  In this case the reduction of the energy occurs along with a comparatively large oscillation of cavity energy.  This oscillation can be quite different for different pulse shapes. \label{energy-fig}}
	\end{figure}

	 We take current elements oriented along \(\boldsymbol{e}_{z}\), and \((x_{1},y_{1},z_{1})=(L-x_{0},L-y_{0},L-z_{0})\).  For concreteness the initial pulse is taken to be a Gaussian centred around \(\omega_{0}\) although our results are not restricted to such a pulse shape
	\begin{equation}
		\mathcal{J}_{0}(\omega)=\frac{J}{\sigma}\sqrt{\frac{\pi}{2}}\left[\e^{-\frac{1}{2\sigma^{2}}\left(\omega-\omega_{0}\right)^{2}}+\e^{-\frac{1}{2\sigma^{2}}\left(\omega+\omega_{0}\right)^{2}}\right]\label{j0}.
	\end{equation}
	The constant \(J\) determines the peak magnitude of the current pulse, and \(\sigma\) the duration.  The sum of two terms in the square brackets ensures the symmetry \(\mathcal{J}_{0}(\omega)=\mathcal{J}_{0}^{\star}(-\omega)\), which leads to a real value for the current in the time domain.  The relationship between the two current pulses (\ref{current-2}) is
	\begin{equation}
		\mathcal{J}_{0}(\omega_{n,m,p})=-(-1)^{n+m+p}\mathcal{J}_{1}(\omega_{n,m,p})\label{current-condition}
	\end{equation}
	where the factor of \((-1)^{n+m+p}\) arises from the even/odd parity of the eigenmodes (this assumes \(\boldsymbol{x}_{0}\neq\boldsymbol{x}_{1}\)).  At first sight it seems that (\ref{current-condition}) cannot be easily fulfilled, because the current is evaluated at the eigenfrequencies \(\omega_{n,m,p}\) which depend on the sum of the \emph{squares} of \(n\), \(m\) and \(p\) rather than their linear sum.  However, it is a property of sets of integers that if their sum is even (odd), then the sum of their squares is also even (odd).  Therefore we can fulfil (\ref{current-condition}) with,
	\begin{multline}
		\mathcal{J}_{1}(\omega)=-\frac{J}{\sigma}\sqrt{\frac{\pi}{2}}\bigg[\e^{-\frac{1}{2\sigma^{2}}\left(\omega-\omega_{0}\right)^{2}}\e^{\frac{\ii (2q+1) L^{2}\omega^{2}}{c^{2}\pi}}\\
		+\e^{-\frac{1}{2\sigma^{2}}\left(\omega+\omega_{0}\right)^{2}}\e^{-\frac{\ii (2q+1) L^{2}\omega^{2}}{c^{2}\pi}}\bigg]\label{box-current}
	\end{multline}
where \(q\) is an integer we are free to choose, according to the desired time delay between the initial pulse and the absorbing one.  An expression for the delay between the two pulses can be found from the expansion of the phase of the first term in (\ref{box-current}) around \(\omega_{0}\)
\begin{multline}
	\frac{(2q+1) L^{2}\omega^{2}}{c^{2}\pi}=\frac{(2q+1) L^{2}\omega_{0}^{2}}{c^{2}\pi}+\frac{2(2q+1) L^{2}\omega_{0}}{c^{2}\pi}(\omega-\omega_{0})\\
	+\frac{2(2q+1) L^{2}}{c^{2}\pi}(\omega-\omega_{0})^{2}.
\end{multline}
The term linear in \(\omega\) corresponds to a time delay between the centre of the first pulse and the centre of the second, \(t_{1}-t_{0}=2(2q+1)L^{2}\omega_{0}/c^{2}\pi\), while the quadratic variation of the phase changes the shape of the second pulse relative to the first.  Figure~\ref{set-up-box}b shows the two pulses for different choices of \(q\).  Although a Gaussian pulse shape does not strictly have a finite duration---as assumed in (\ref{jw})---one may truncate the infinite tails at some point when the amplitude of the current is arbitrarily small and then apply the argument of section~\ref{abs-sec}.  Figure~\ref{energy-fig} shows the cycle--averaged energy in the cavity as a function of time, computed from (\ref{cavity-energy}), for the case \(q=1\) (for details, see appendix~\ref{apa}).
%
%
	%
	%
	\subsection{Thin cylindrical cavity}
	\par
\begin{figure}[h!]
		\includegraphics[width=8.5cm]{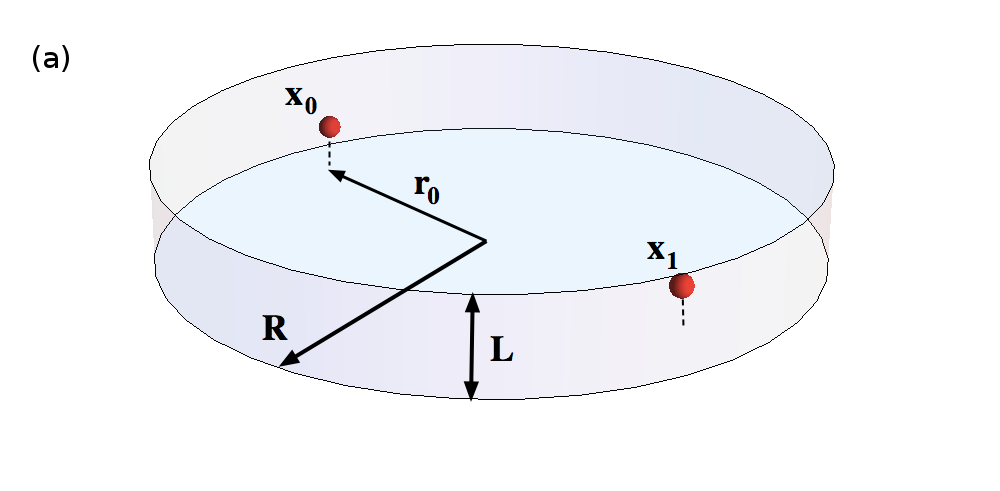}
		\includegraphics[width=8.5cm]{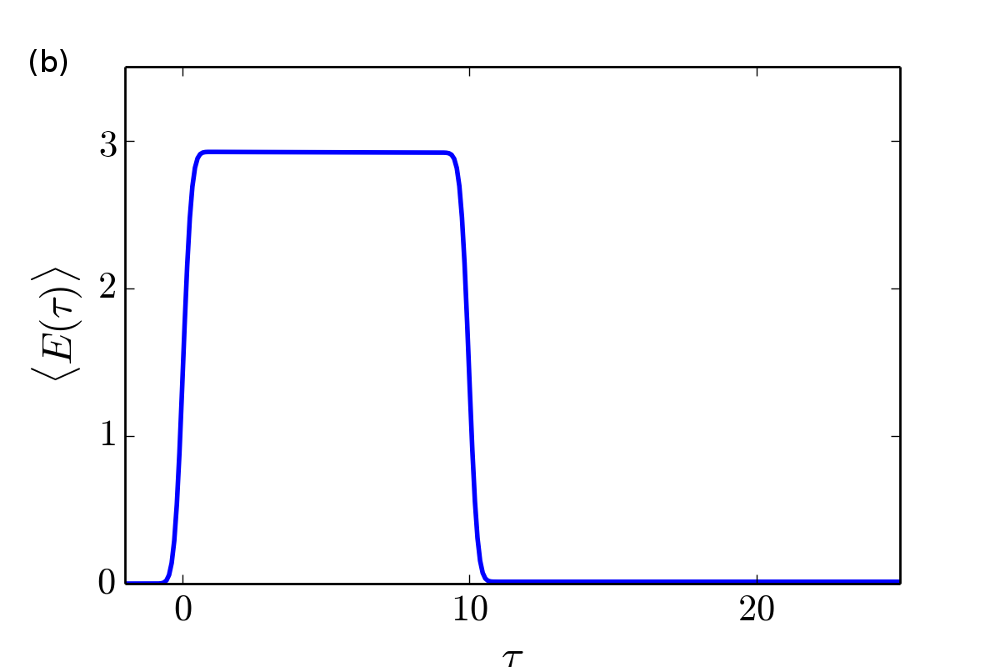}
		\caption{(a) Two sources at opposite positions within a thin cylindrical cavity emit sequential pulses.\label{cylinder-fig} (b) Energy in the cavity as a function of time, scaled in units of \(\mu_{0}J^{2}/L\).  The expression for \(\langle E(\tau)\rangle=L\langle\mathscr{E}(\tau)\rangle/\mu_{0}J^{2}\) is given in appendix~\ref{apb}.  In this case \(r_{0}=0.01R\), \(\sigma=3 c/R\) and \(\omega_{0}=20 c/R\).}
	\end{figure}

	As a second example we consider a thin cylindrical cavity of thickness \(L\), where \(L\ll 2\pi c/\omega\) for the frequencies of interest.  In this regime, the thickness is negligible with respect to the wavelength, which makes the waves effectively two-dimensional and forces the electric field to be polarized along the cylinder axis. The normalized eigenmodes are then given by
	\begin{equation}
		\boldsymbol{E}_{n,l}(r,\theta)=\sqrt{\frac{2}{V [J_{l+1}(\omega_{n,l} R/c)]^2}}\,\boldsymbol{e}_{z}J_{l}(\omega_{n,l} r/c)\cos(l\theta)\label{cylindrical-modes}
	\end{equation}
	where \(V=\pi R^{2}L\) is the volume of the cavity, \(J_{l}\) is the Bessel function of the first kind and the eigenfrequencies are given by
	\[
		\omega_{n,l}=\frac{c}{R}j_{l,n}
	\]
	where \(j_{l,n}\) is the position of the \(n\)th zero of the \(l\)th order Bessel function of the first kind~\cite{dlmf}.  We again choose the absorption point opposite to the emission point (\(r_{1}=r_{0}\), \(\theta_{1}=\theta_{0}+\pi\); the \(z_{0,1}\) coordinates don't matter because in this regime the modes are \(z\) independent), as shown in figure~\ref{cylinder-fig}, and from (\ref{current-2}) we find that the current pulses must be related by
	\begin{equation}
		\mathcal{J}_{0}(\omega_{n,l})=-(-1)^{l}\mathcal{J}_{1}(\omega_{n,l})\label{condition-2}.
	\end{equation}
	%
	%
	\par
	For the pulse emitted from \(\boldsymbol{x}_{1}\) to absorb the pulse emitted from \(\boldsymbol{x}_{0}\), we have to be able to write \(-(-1)^{l}\mathcal{J}_{l}(\omega_{n,l})\) as a function of \(\omega\) that interpolates \(\mathcal{J}_{0}(\omega)\) at the eigenfrequencies.  If we attempt to make (\ref{condition-2}) hold for all \(n\) and \(l\) we find that, due to the spacing of the zeros of Bessel functions of different orders, the required function generally oscillates wildly and irregularly as \(\omega\) increases.  However, when the source and emission point are close to the centre of the cavity their coupling to the higher \(l\) eigenmodes is much reduced, which makes the secondary pulse much better behaved.
	\par
	The simplest case is where both \(\boldsymbol{x}_{0}\) and \(\boldsymbol{x}_{1}\) are close to the centre of the cavity, and the initial pulse is centred around a frequency \(\omega R/c\gg1\).  For large \(n\) and \(n\gg l\), the zeros of the Bessel functions are approximately,
	\begin{equation}
		j_{l,n}\sim\left(n+\frac{l}{2}-\frac{1}{4}\right)\pi\label{approx-zeros}
	\end{equation}
	so that we can fulfil (\ref{condition-2}) with,
	\begin{multline}
		\mathcal{J}_{1}(\omega)=-\frac{J}{\sigma}\sqrt{\frac{\pi}{2}}\e^{2i(2q+1)\left(\frac{\omega R}{c}+\frac{\pi}{4}\text{sign}(\omega)\right)}\\
		\times\bigg[\e^{-\frac{1}{2\sigma^{2}}\left(\omega-\omega_{0}\right)^{2}}
		+\e^{-\frac{1}{2\sigma^{2}}\left(\omega+\omega_{0}\right)^{2}}\bigg]\label{x1-current}
	\end{multline}
	where \(q\) is again an integer that we are free to choose, and determines the delay of the second pulse.  In this approximation the second pulse is of the same shape as the initial pulse but delayed by \(t_{1}-t_{0}=2(2q+1)R/c\), which is the time taken for light to travel to the walls of the cavity and back again an odd number of times.  The approximation gets worse with increasing \(q\) due to the fact that the exponential in (\ref{x1-current}) was found from (\ref{approx-zeros}), and the error in this approximation is multiplied by \(2q+1\) within the exponent.  Figure~\ref{cylinder-fig} shows the cycle--averaged energy in the cavity for the case \(q=2\).
	%
	%
	\section{Experimental results}
	\par
	In order to verify our results experimentally we took a metallic cubic cavity with sides of \(L=15\,\text{cm}\) so that the eigenfrequencies occur in the microwave regime: \(\omega_{n,m,p}=2\pi\sqrt{n^{2}+m^{2}+p^{2}}\,\times10^{9}\,\text{rad}\,\text{s}^{-1}\).  Two small antennas (a few millimeters of exposed coaxial cable) were inserted through two holes drilled into the middle of opposite sides of the box, one acting as a probe of the field in the cavity, and one acting as both source and drain.  The probe antenna was roughly half the length of the source.  The theory discussed in section~\ref{cavsec} applies to this situation, but for the simpler case of \(\boldsymbol{x}_{1}=\boldsymbol{x}_{0}\).  In this case one must remove the factor of \((-1)^{n+m+p}\) from (\ref{current-condition}), and the phase factor in (\ref{box-current}) becomes \(\exp(2\ii qL^{2}\omega^{2}/\pi c^{2})\).
	
\begin{figure}[h!]
		\includegraphics[width=8cm]{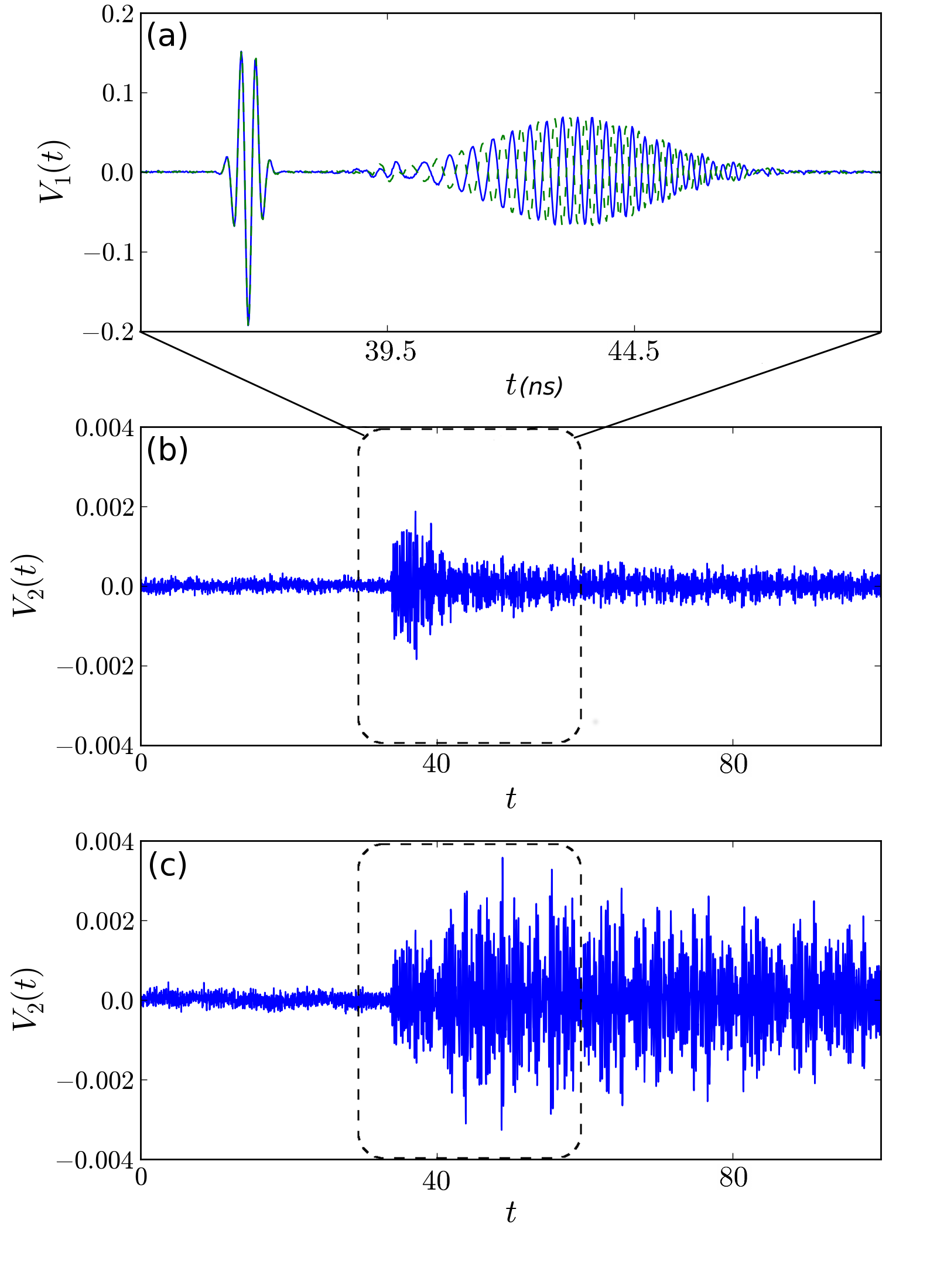}
		\caption{(a) Measured output from the AWG.  The \(x\) axis is in units of nano--seconds and the \(y\) axis is in units of volts.  The solid line shows the voltage sent into the cavity for the case shown in panel b, and the dashed line is for the case shown in panel c.  (b)  Measured field inside the cavity for the case when the second pulse ought to reduce the cavity field to zero.  The dashed boxes show the interval of time given in panel (a).  (c)  Measured field inside the cavity, with the second pulse multiplied by \(-1\).\label{exp-fig}}
	\end{figure}

	The source/drain antenna was attached to an arbitrary waveform generator (AWG) (Tektronix AWG70000 series), and the probe was attached to an oscilloscope (Tektronix MSO70000 series) to monitor the field inside the cavity.  From the analytic expressions for the pulse shapes necessary to absorb the field we generated a time series of voltage values which were imported into the AWG to generate the pulses.  The AWG outputs the desired time domain signal, then starts again at the beginning---we included a long gap of zero output (\(>1\,\mu\text{s}\)) to let the field in the cavity decay to a small value before emitting the pulses again.  Two cases were compared to one another: (i) the case when the second pulse ought to reduce the field in the cavity to zero; and (ii) the same as (i), but with the second pulse multiplied by \(-1\).  Figure~\ref{exp-fig}a shows the output from the AWG for these two cases (solid and dashed lines respectively).  Figures~\ref{exp-fig}b and c show the measured field inside the cavity as a function of time.  In the case of figure~\ref{exp-fig}b, the second pulse ought to reduce the cavity field to zero.  The field is clearly reduced by the second pulse, although it does not return to the pre-pulse level (note that there is always some residual noise in the cavity).  One possible reason for this discrepancy is that we have not included the lifetime of the eigenmodes in our theory (measured cavity lifetime \(\sim 1\,\mu\text{s}\)).  It is worth comparing figure~\ref{exp-fig}b to figure~\ref{exp-fig}c which differs only in the sign of the second pulse, and shows a much larger cavity field; in this case the second pulse, instead of extracting the cavity field, adds the same field once more, multiplying the original field by the factor of two and the energy by the factor of four.  
	%
	%
	%
	%
	\section{Conclusions}
	
	We have shown that if a pulse is launched into an empty cavity, then it may be completely absorbed at a later time through launching a second pulse.  The shape of the second pulse is a function of the first one and the eigenfrequencies of the cavity.  In particular, a cubic cavity and a cylindrical one have a particularly simple relationship between the two pulses, and we have verified by direct computation that the cavity energy is reduced to zero.  In these cases the shape of the absorption pulse can be relatively simple and of a duration that is comparable to the initial pulse.  This is thanks to the symmetry of the cavity.  For a general cavity shape without such symmetry, the absorption pulse is often incomparably longer than the emission pulse~\cite{tyc2012}.
	
	From figure~\ref{energy-fig} it is clear that the way in which the field energy is reduced to zero can be quite unlike the time reversal operation discussed by de Rosny and Fink~\cite{derosny2002}.  We also experimentally demonstrated this effect in a cubic cavity, showing that through a judicious choice of the second pulse the field in the cavity could be either amplified or diminished, in broad agreement with our theory.  
	
	\acknowledgements
	SARH and RNF thank the University of Birmingham for hospitality and for kindly letting us use their equipment, and Tektronix for loaning us the 70000 series AWG and oscilloscope.  We also thank J. R. Sambles, A. P Hibbins, and I. R. Hooper for useful discussions, and we acknowledge financial support from the EPSRC under Program Grant EP/I034548/1.  TT acknowledges financial support from grant no. P201/12/G028 of the Czech Science Foundation.\\
%
%
	\appendix
	
	\section{Evaluating the energy in a cubic cavity\label{apa}}
	\par
	To compute the energy we used the expressions for the two kinds of modes within the cavity \(\boldsymbol{E}^{(1,2)}\) given by
\begin{align}	\boldsymbol{E}^{(1)}_{n,m,p}&=\left(\frac{L}{\pi}\right)\boldsymbol{e}_{z}
\boldsymbol{\times}\boldsymbol{\nabla}_{\parallel}
\psi^{(1)}_{n,m,p}\nonumber\\
\boldsymbol{E}^{(2)}_{n,m,p}&=\left[\boldsymbol{e}_{z}\left(n^{2}+m^{2}\right)
+\left(\frac{L}{\pi}\right)^{2}\boldsymbol{\nabla}_{\parallel}\frac{\partial}{\partial z}\right]\psi^{(2)}_{n,m,p}\label{cavity-modes}
\end{align}
where \(\boldsymbol{\nabla}_{\parallel}=\boldsymbol{e}_{x}\frac{\partial}{\partial x}+\boldsymbol{e}_{y}\frac{\partial}{\partial y}\), and
\begin{align}
	\psi^{(1)}_{n,m,p}&=N^{(1)}_{n,m,p}\cos\left(\frac{n\pi x}{L}\right)\cos\left(\frac{m\pi y}{L}\right)\sin\left(\frac{p \pi z}{L}\right)\nonumber\\
	\psi^{(2)}_{n,m,p}&=N^{(2)}_{n,m,p}\sin\left(\frac{n\pi x}{L}\right)\sin\left(\frac{m\pi y}{L}\right)\cos\left(\frac{p \pi z}{L}\right).
\end{align}
In (\ref{cavity-modes}), one of the integers \(n,m,p\) may be zero.  This is distinct from the case of scalar waves, where all of the integers must be greater than zero~\cite{tyc2012}.  This particular representation of the modes within the cavity (\ref{cavity-modes}) can be obtained from the expressions for the electric field in a rectangular waveguide~\cite{volume8} (propagation axis \(\boldsymbol{e}_{z}\)), applying the boundary condition that the tangential electric field is zero at the ends of the guide \(z=0,L\).  The normalization of the modes is chosen so that the modes satisfy (\ref{orthonormal}):
\begin{align*}
	N^{(1)}_{n,m,p}&=\left(\frac{2}{L}\right)^{3/2}\sqrt{\frac{1}{n^{2}+m^{2}}}\frac{1}{\sqrt{(1+\delta_{n0})(1+\delta_{m0})}}\\
	N^{(2)}_{n,m,p}&=\left(\frac{2}{L}\right)^{3/2}\sqrt{\frac{1}{(n^{2}+m^{2})(n^{2}+m^{2}+p^{2})(1+\delta_{p0})}}.
\end{align*}
When computing the energy (\ref{cavity-energy}), we assume the two sources are oriented along the \(\boldsymbol{e}_{z}\) axis so that they only couple to \(\boldsymbol{E}^{(2)}_{n,m,p}\).  The expansion coefficients (\ref{expansion-coefficient}) then take the form
	\begin{widetext}
	\begin{equation}
		\mathcal{C}_{n,m,p}^{(2)}(\tau)=\frac{2\mu_{0}J\sqrt{L}}{\sqrt{\pi}s}\sqrt{\frac{n^{2}+m^{2}}{(n^{2}+m^{2}+p^{2})(1+\delta_{p0})}}\sin\left(\frac{n\pi x_{0}}{L}\right)\sin\left(\frac{m\pi y_{0}}{L}\right)\cos\left(\frac{p \pi z_{0}}{L}\right)\text{Re}\left[I_{n,m,p}(\tau)\right]\label{exp-1}
	\end{equation}
	where
	\begin{equation}
		I_{n,m,p}(\tau)=\int_{-\infty}^{\infty}d\xi\frac{\left[1-(-1)^{n+m+p}\e^{\frac{\\i (2q+1) \xi^{2}}{\pi}}\right]}{\pi^{2}(n^{2}+m^{2}+p^{2})-(\xi+\\i\eta)^{2}}\e^{-\frac{(\xi-\xi_{0})^{2}}{2s^{2}}}\e^{-\\i\xi\tau}\label{Iint}
	\end{equation}
	and \(\mathcal{J}_{0}\) is given by the same expression as listed in figure~\ref{set-up-box}, with \(\sigma=s c/L\) and \(\omega_{0}=\xi_{0} c/L\).  Inserting (\ref{exp-1}) into (\ref{cavity-energy}), we obtain the following expression for the time averaged energy in the cavity
	\begin{equation}
		\frac{L\langle\mathscr{E}_{R}(\tau)\rangle}{\mu_{0}J^{2}}=\frac{1}{\pi s^{2}}\sum_{n,m,\in\text{odd}}\sum_{p=0}^{\infty}\frac{n^{2}+m^{2}}{(n^{2}+m^{2}+p^{2})(1+\delta_{p0})}\left[\left|\frac{d I_{n,m,p}(\tau)}{d\tau}\right|^{2}+\pi^{2}\left(n^{2}+m^{2}+p^{2}\right)\left|I_{n,m,p}(\tau)\right|^{2}\right]\label{energy-series}
	\end{equation}
	which has been scaled by the characteristic energy \(\mu_{0}J^{2}/L\).  The position \(\boldsymbol{x}_{0}\) has been chosen as \((L/2,L/2,0)\), leading to the sum over only odd integers \(n\) and \(m\).  Equation (\ref{energy-series}) was numerically evaluated for the case \(q=1\), and the arbitrary value \(\eta=10^{-4}\) in order to obtain figure~\ref{energy-fig}.
	\end{widetext}
	
	\section{Evaluating the energy in a cylindrical cavity\label{apb}}
	
	From the expression for the modes in the cavity given by (\ref{cylindrical-modes}), the expansion coefficients (\ref{expansion-coefficient}) are given by
	\[
		\mathcal{C}_{n,l}(\tau)=\frac{\mu_{0}JR}{\pi s\sqrt{L}}\frac{J_{l}(j_{l,n}r_{0}/R)}{\left| J_{l+1}(j_{l,n})\right|}\text{Re}\left[I_{n,l}(\tau)\right]
	\]
	where \(\theta_{0}=0\) and
	\begin{multline}
		I_{n,l}(\tau)=\int_{-\infty}^{\infty}d\xi
		\frac{\left[1-(-1)^{l}\e^{2{\rm i} (2q+1)(\xi+\frac{\pi}{4}\text{sign}(\xi))}\right]}{j_{l,n}^{2}-(\xi+{\rm i}\eta)^{2}}\\
		\times \e^{-\frac{(\xi-\xi_{0})^{2}}{2s^{2}}}\e^{-{\rm i} x\tau}
	\end{multline}
	In this case the time variable \(\tau=c t/R\) is scaled by the time taken to get from the centre of the cavity to the edge.  This leads to the following expression for the cavity energy
	\begin{multline}
		\frac{L\langle\mathscr{E}_{R}(\tau)\rangle}{\mu_{0}J^{2}}=\frac{1}{4\pi^{2}s^{2}}\sum_{n,l}\left(\frac{J_{l}(j_{l,n}r_{0}/R)}{J_{l+1}(j_{l,n})}\right)^{2}\\
		\times\left[\left|\frac{dI_{n,l}(\tau)}{d\tau}\right|^{2}+x_{n,l}^{2}\left|I_{n,l}(\tau)\right|^{2}\right]
	\end{multline}
	which is plotted in figure~\ref{cylinder-fig}.
	
\end{document}